# How persuadee's psychological states and traits shape digital persuasion: Lessons learnt from mobile burglary prevention encounters

Mateusz Dolata
University of Zurich
dolata@ifi.uzh.ch

Robert O. Briggs
San Diego State University
rbriggs@sdsu.edu

Gerhard Schwabe
University of Zurich
schwabe@ifi.uzh.ch

**Abstract**

*Persuasion can be a complex process. Persuaders may need to use a high degree of sensitivity to understand a persuadee's states, traits, and values. They must navigate the nuanced field of human interaction. Research on persuasive systems often overlooks the delicate nature of persuasion, favoring "one-size-fits-all" approaches and risking the alienation of certain users. This study examines the considerations made by professional burglary prevention advisors when persuading clients to enhance their home security. It illustrates how advisors adapt their approaches based on each advisee's states and traits. Specifically, the study reveals how advisors deviate from intended and technologically supported practices to accommodate the individual attributes of their advisees. It identifies multiple advisee-specific aspects likely to moderate the effectiveness of persuasive efforts and suggests strategies for addressing these differences. These findings are relevant for designing personalized persuasive systems that rely on conversational modes of persuasion.*

## 1. Introduction

Effective persuasion is essential for successful advice-giving (Dolata et al., 2016; Dolata & Schwabe, 2018). However, even when facing high-stakes problems, many persuadees do not act on advice (Bonaccio & Dalal, 2006). For example, many patients disregard doctors' orders (Seiders et al., 2015), clients ignore financial advice (Bradbury et al., 2015), and citizens neglect to enhance home security despite recommendations from experienced police advisors (Comes & Schwabe, 2016b). Unpersuaded advisees remain in difficult situations that can grow worse over time causing problems for themselves and the society.

Persuasion is the deliberate effort to shape, reinforce, or change the attitudes, behaviors, or beliefs of a person without deceit or coercion, under conditions where persuadee is free to accept or reject the recommendations (O'Keefe, 2016). A persuader can motivate change, enhance someone's ability to change, prompt action towards change, or solidify a decision or behavior once a change has been made (Cialdini, 2007).

Research shows that persuasive technologies (PTs) can increase the likelihood of individuals following expert advice. Persuasive technologies are tools intended to be used to increase the likelihood that a persuadee will follow through on a recommendation (Fogg, 2009; Stibe, 2015). For example, studies have shown that mobile PTs can be associated with positive behavioral changes, such as improved exercise and eating habits, increased water conservation, and better adherence to infection prevention measures (Oyibo & Morita, 2021). A significant portion of scholarly research indicates that PTs can lead to varying degrees of positive change (Hamari et al., 2014; Orji & Moffatt, 2018).

Persuasive technologies, though, are not yet consistently effective. Studies show that their impact can be unpredictable, sometimes yielding mixed or negligible results, and in certain cases, even increasing resistance (Aldenaini et al., 2020). The effectiveness of a specific PT fluctuates based on various factors such as personality, gender, age, and other characteristics of the individual being persuaded (Alkış & Taşkaya Temizel, 2015; Halttu & Oinas-Kukkonen, 2022; Kaptein et al., 2015).

Research shows that persuasion relies on delivering the right message, at the right time, and in the right way (Kaptein et al., 2015). However, there is still no clear consensus in the literature on how to devise a "right" message, timing, or approach for a given persuasion attempt. To address this gap, investigation is needed to understand the communication patterns, behaviors, and conditions linked to high and low compliance. With persuasive applications rapidly expanding across mobile and AI platforms, it is crucial to investigate how persuasion efforts can be individualized and personalized.

This paper presents a retrospective analysis of research archives compiled over eight years during a three-iteration Design Science Research (DSR) program called Smart Protector (SP). SP focused on augmenting burglary prevention services offered by the police to enhance the overall quality and effectiveness of the advisory service. It ended in 2020. Testing in both experimental and real-world settings confirmed the system's potential to significantly enhance the advisory services' quality. The system was later used in burglary prevention advisor' daily practice.

Despite promising results and practical success, the researchers observed inconsistencies in how police officers utilized the system across different advisory sessions. It was unclear why advisors deviated from the intended and trained behaviors. Our preliminary analysis suggested that differences among advisees could be a source of the deviations, prompting further investigation. It was driven by the following research question:

*RQ. How and why do perceived characteristics of persuadees prompt persuaders to vary their behavior?*

To explore this question, we conducted a retrospective analysis of research material collected during the SP project in the spirit of design archeology (Chandra Kruse et al., 2019). The purpose was to discover utterances, behaviors, situations, issues, and events that were associated with advisors' following or deviating from an intended practice. Our analysis revealed sophisticated considerations advisors employ during persuasive efforts. Notably, advisors often attend to the advisees' mental states and personality traits, incorporating these into persuasion to achieve best impact.

Specific findings from this research explain the seemingly idiosyncratic use of the SP system. It reveals how practitioners implement persuasive techniques and what concerns they intuitively make. This understanding offers a two-fold benefit: it provides guidance for developing systems that better support advisors, and informs the design of more user-centered persuasive technologies. Insights from this study hold particular relevance for designing and developing persuasive tools that utilize natural language communication. This is especially applicable to the field of LLM-powered chatbots, which are increasingly prevalent in applications related, for instance, to digital healthcare and well-being.

## 2. Background

### 2.1. Advisory Services

Advisory services represent a unique form of collaboration between an advisor and advisee. From a service-science perspective, this interaction is a high-touch point, intensifying the relationship between provider and customer (Jungermann & Fischer, 2005). Conversation studies view this as institutional talk, where participants embody their organizational identities through language, materials, and behavior (Dolata et al., 2019; Dolata & Schwabe, 2017; Drew & Heritage, 1992). Collaboration support research sees it as collaboration between individuals with potentially divergent goals and knowledge (Heinrich et al., 2014; Heyman & Artman, 2015). Finally, many studies emphasize the persuasive element of advice-giving, where the pursuit of the best solution intertwines with the advisor's efforts to motivate the advisee's follow-through. This perspective is most prevalent in doctor-patient encounters to support patient's adherence to treatment (Rubinelli, 2013), but has also been applied to crime prevention (Dolata & Schwabe, 2018). The perspectives are tightly intertwined. For instance, an advisor might leverage their institutional identity to increase persuasiveness.

Diverse perspectives on advisory services underscore the crucial role of nuanced interpersonal dynamics, particularly in infrequent collaborations between individuals unfamiliar with one another (Heinrich et al., 2014; Jungermann & Fischer, 2005). This inherent complexity leads advisory encounters to rely heavily on pre-existing stereotypes held by both advisees and advisors regarding expected behaviors and outcomes (Dolata et al., 2019; Jungermann & Fischer, 2005; Svinhufvud & Vehviläinen, 2013). At the same time, research demonstrates that personalized and individualized advisory services enhance client satisfaction (Saxe & Weitz, 1982). Thus, advisors navigate a delicate balance between fulfilling stereotypical expectations and meeting the demand for empathy, active listening, and personalized approach (Dolata et al., 2019). How they successfully manage this tension remains an open question.

Burglary prevention can be a sensitive area to navigate due to its delicate domain. When discussing this topic, individuals might feel vulnerable as they are inviting a police officer to inspect their home, thereby exposing their private sphere. Also, they are being asked to reassess their approach to security, a fundamental human need. The advice can challenge their sense of safety and question deeply ingrained habits. Past research indicates that persuasive systems can effectively support burglary prevention advisors in persuasion efforts (Comes & Schwabe, 2016a; Dolata et al., 2016)

### 2.2. Persuasive Systems

Persuasive systems have been traditionally defined as *"computerized software or information systems designed to reinforce, change or shape attitudes or behaviors or both without using coercion or deception"* (Oinas-Kukkonen & Harjumaa, 2009). This category initially encompassed technologies designed for (1) computer-human influence, (2) computer-mediated human-human influence, and (3) computer-moderated influence (Dolata et al., 2016; Oinas-Kukkonen & Harjumaa, 2009; Stibe, 2015). Persuasive systems in the first category operate on the premise that technology can function as a social agent, thereby influencing individual behavior. When technology acts as a mediator (e.g., in blogs, forums, and social networks), individual behavior and attitudes are subject to social influence through user-generated content disseminated by the technology.

Systems in the third category leverage information about the behavior of others to influence a user's behavior or attitudes, promoting behavior-based social influence (Stibe, 2015). While the field continues to evolve, its core focus remains on utilizing technologies to modify the attitudes and behaviors of system users (Aldenaini et al., 2020; Alslaity et al., 2024).

A distinct form of persuasive systems arises when humans, utilizing technology as a tool, assume a significant role in the persuasive process (Dolata et al., 2016). In such sociotechnical systems, while a human persuader holds primary responsibility for persuasion, technology plays a supporting role, offering tools that the persuader can leverage as needed. This approach is evident in various domains, including burglary prevention advice, healthcare counseling, and other advisory services (Staehelin et al., 2023). Despite its practical significance, this perspective on persuasive systems remains relatively unexplored.

Blending technology and human persuasion aims to leverage the innate human abilities. The research deals with how persuasion needs to be encoded and what psychological processes are involved in decoding it. Existing research on human-human persuasion can be categorized by its focus (O'Keefe, 2016 provides a useful summary): Some research seeks to establish general principles for crafting effective persuasive messages, yielding practical guidance and techniques. Others examine communicator characteristics, such as credibility or attractiveness, and their impact on persuasion's effectiveness. A third area concentrates on the recipient, emphasizing the influence of factors like mood and initial reactions on the persuasive process. Cultural beliefs can significantly influence how convincing arguments are perceived (Davis et al., 2018). Other factors related to the persuadee such as their readiness to change and their level of resistance (reactance), are also likely to impact how credible they find a persuasive message (Beutler et al., 2018; Krebs et al., 2018).. Adaptation to the persuadee was shown to improve persuasion efforts, e.g., in the context of psychotherapy (Vaz & Sousa, 2021). While additional factors, like counterfactual thinking, have also been explored, a consensus remains that further investigation is needed to grasp the recipient's role in persuasion. Specifically, more research is needed to understand which recipient characteristics should be considered when attempting persuasion (O'Keefe, 2016; Stavraki et al., 2021; Vaz & Sousa, 2021).

Research in persuasive systems is increasingly focused on adapting to persuadee variables. Moving beyond generic approaches, researchers are designing systems that personalize persuasion (Oyebode et al., 2024). This shift towards personalization is particularly evident in health and wellness, where studies explore how user characteristics (e.g., age, gender, personality) impact the effectiveness of behavior change interventions (Alslaity et al., 2024). For example, persuasive systems have been developed to support individuals at different stages of behavior change (Oyebode et al., 2024). Tailoring persuasive systems to individual personalities and behavior stages has proven to be more effective in motivating healthy behaviors (Alslaity et al., 2024). Yet, while existing studies highlight relevant user variables, they primarily focus on computer-human influence. To advance this area, we argue that there is much to be learned from human persuaders, particularly regarding the effective tailoring of persuasion to more nuanced user aspects.

A deeper understanding of persuadee variables can fuel the next generation of persuasive technologies. While the first generation focused on broadly effective persuasion approaches, limited by technology's personalization capabilities, Generative AI (GenAI) allows for easy and more precise tailoring persuasive interventions to individual user profiles (Matz et al., 2024; Wu et al., 2024). This aligns with calls for recognizing individual characteristics to persuasive techniques (Matz et al., 2024). However, it's unclear which persuadee aspects are crucial for designing GenAI-based persuasive systems. Should they consider a user's openness, emotional state, demographics, or other factors? Learning the strategies employed by human persuaders could provide insights into relevant persuadee attributes. Also, this could illuminate how tools can be utilized differently based on advisors' perceptions of these attributes.

## 3. The Smart Protector

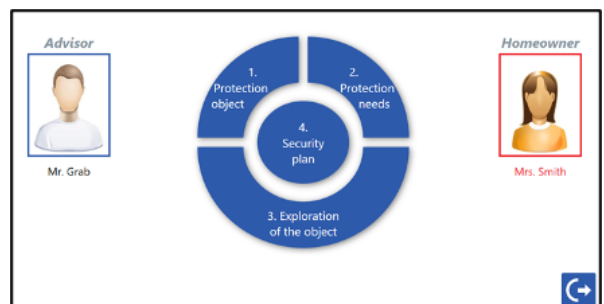

**Figure 1. The advisors process navigation screen.**

The SP system was developed in a user-centered process. Beginning in 2012, the requirements of advisors were gathered through observations of key users' daily work, in workshops held in 2013, and again during the design of working prototypes. Throughout 2013-2016, these prototypes were tested in simulated experiments with various BP advisors from Switzerland and Germany. Since then, the system has been released and remains in use by police advisors.

SP is a tablet-based system designed for advisors conducting advisory encounters at advisees' homes. This portable, mobile system offered several key fea-tures for generating individualized documentation, visually demonstrating home security issues and their solutions, communicating risks using multimedia, identifying and recording advisee needs, and prioritizing post-visit action steps for the advisee.

SP guides users through a structured advisory process (Figure 1), with each step supported by specific functionalities. First, both advisor and advisee input information about the house (*protection object*), including basic data and a photograph. This personalized approach adds immediacy to the service and documentation. Next, users identify and discuss advisees' most pressing *protection needs*. A predefined set of common needs prompts a discussion about safety fears and feelings – primary motivators for crime prevention. Participants then *explore the object* to pinpoint vulnerabilities (Figure 2). For each vulnerability, the advisor can: classify the problem (steps 1 & 2); select or add a description (3); show a video or graphic explaining the problem and related risks (4); photograph the specific flaw (5); and propose a solution (6). Videos improves risk awareness, photos aid personalized documentation, and linking solutions enhances the advisee's self-efficacy. Finally, the advisor and advisee collaborate on a prioritized *security plan* with actionable recommendations, further bolstering the advisee's self-efficacy. For a detailed description of these functionalities, the relevant literature, and their measured effects please refer to the following articles: Comes & Schwabe, 2016a, Comes & Schwabe, 2016b, Dolata et al. 2016, Dolata & Schwabe 2018.

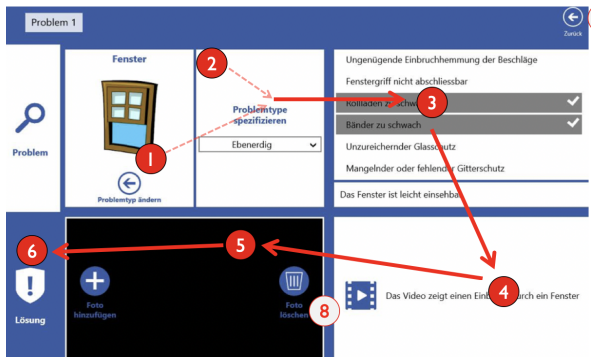

**Figure 2. The problem specification screen and suggested (though not enforced) process steps. The picture comes from the actual training material provided to advisors.**

Advisors received dedicated training on the system, which included feature explanations, practical grounding from advisors involved in the design process, trial advisory services with peers, and several sessions with figurants. Despite this, post-rollout, advisors deviated from the intended practices in an inconsistent manner as indicated in shadowing sessions and interviews. For example, an advisor might utilize a specific function in one service but not another. This contradicted the system's goal of reliably improving follow-through by enabling persuasive behaviors in all advisory services. Intrigued by these findings, researchers conducted a retrospective analysis of project data, drawing inspiration from design archaeology, which advocates for examining past design decisions and outcomes to glean insights (Chandra Kruse et al., 2019), which – in our case – are primarily focused on the impact of the design artifact in the real world and users' reactions to it.

## 4. Methodology

This paper analyzes data gathered during the SP project. Three researchers reviewed the archived research material from the SP project to discover incidents, utterances, and actions of advisors and advisees that could inform the research questions. The archived research material included 136 audio-video recordings of advisory services ranging from 45 to 90 minutes including 48 fully transcribed recordings; 189 audio-recordings of after-action interviews with advisees and advisors including 64 fully transcribed ones; materials from 24 persuaders workshops, including transcribed audio-recordings of 9 workshops; 174 pages of handwritten notes of researchers' field observations; more than 200 brochures and copies of police's materials handed to the researchers and/or to the advisees; anonymized reports from the advisory encounters; and photos of houses and security flaws from 550 advisory encounters conducted with the persuasive technology upon organizational implementation; seven published papers reporting on results from various stages of the projects (published between 2014 and 2018); nine bachelor's or master's theses submitted between 2014 and 2020.

The researchers reviewed the archives seeking recurring instances of utterances, behaviors and events that were associated with advisors' deviating from the intended persuasion behaviors. Researchers identified a collection of critical incidents and utterances that provided insights about why advisors and advisees evidenced a strong preference for unstructured advisory processes and seemingly idiosyncratic practices occurring in various situations. The researchers compiled these discoveries into a repository of critical incidents accompanied by comments. Then they used open coding (Saldana 2009) to organize critical incidents according to the dimensions that advisors considered when explaining or arguing for a certain behavior as the most appropriate one or which appeared as relevant from their conduct. These provided new understandings of the persuader-persuadee relationship.

Many of the discovered insights derived from this exploration seem obvious in retrospect, but they were not self-evident when the research began. They emerged layer-by-layer as the retrospective analysis unfolded.

The following section presents the concerns that drive advisors' preference for certain behaviors when interacting with specific clients or in specific circumstances.

## 5. Results

The analysis revealed diverse factors advisors consider when engaging in persuasion as indicated in numerous statements and critical incidents. Introducing new technology and advisory approaches through the SP project prompted critical self-reflection among advisors. It forced conscious and unconscious decisions about integrating new tools into existing workflows. Data analysis revealed behavioral variations in advisory approaches, raising questions about their root causes.

To understand these differences, researchers analyzed interviews conducted towards the end of the project and conversations recorded during 2016 shadowing sessions. During these sessions, researchers observed each advisor's daily work for one or two days, focusing on their use of the SP system. This analysis revealed numerous instances where advisors reflected on factors influencing their advising approach Statements are coded for clarity: "I" denotes interviews, "S" represents shadowing session conversations, and "Bxx" refers to specific advisor IDs. Advisor statements and researcher observations are categorized as relating to either advisee's temporary states or their traits. The categorization used below emerged in a bottom-up manner.

### 5.1. Perceived States

The data indicates that advisors need to react to various internal states of the advisees. We define such a state as a temporal property of someone's mind which might embrace perceptions, experiences, or intentions. States might be conscious or unconscious. The advisors adapt their behaviors based on mental states they attribute to the advisees. In the following we attend to states which repetitively occur in the data set.

***Emotional state:*** One's emotional state is the magnitude, direction, and classification of one's affective arousal. Emotional states can range from intense positive emotion (e.g., joy) to no emotion to intense negative emotion (e.g., wrath).

Advisors often find themselves navigating intense emotions during their sessions. For instance, persons who were victims of burglary in the past might experience such profound negative emotions that they are impossible to conceal, ultimately resulting in an emotional outburst. An advisor explains: *"Many times, you notice that (...) the most important thing is actually almost in the background, that's still exciting for me. I would never say: 'look at the window, in twelve seconds there's one in there (...)'. I wouldn't do that. Because, you just spread fear and terror and the woman may have a completely different need. That's quite crucial for me, they are afraid and you can say, 'hey, bedroom robber, that's fortunately history, maybe once we have a case, but if you see the relations, with 7,000 residential burglaries, maybe once we have a case where there is a confrontation.' I mean, if you take it that way, from that side, they're already happy."* (I-B18). This advisor explains that when they sense anxiety or other negative emotions in their advisees, even if not explicitly expressed, they prioritize positive messaging over negative feedback.

***Self-efficacy state:*** One's self-efficacy state (related to expectancy state, likelihood state) is the degree to which one believes that one will be able to attain a given goal.

The research team noticed that many individuals did not consider themselves able to implement the changes proposed by the advisor because they were perceived as too complex, demanding too much attention and time, contrary to their routines, or too expensive. An advisor comments: *"There are people who are very motivated but do not have enough resources or have concerns about large expenses. And sometimes they're ashamed to admit that. Then it can be useful to provide substitutes with great simplicity. 'Look, you can also achieve high security with Plexiglas or a simple iron bar.' (...) I still bring them [substitutes] into the discussion on my own initiative. And people are happy about that. They then think, 'Okay, I'll save some money here in the basement. I can invest that in the other windows upstairs.'"* (S-B06). This behavior demonstrates how advisors strive to help advisees overcome their limitations and empower them to effect positive change.

***Risk-aversion state:*** One's risk aversion state is the degree to which one is willing at a given moment to expose one's self to potential harm. Risk aversion can range from extreme risk seeking to extreme risk avoidance. One's risk aversion state can vary in response to external stimuli or internal contemplation.

Many advisees declare to do anything needed to secure their house because they do not want to put themselves in any unexpected situation, while others are more prone to taking the risk claiming, e.g., that there is nothing of particular value in their homes. The ones want to be as certain as possible, while the others can deal with uncertainty to a certain degree. An advisor refers to it as follows: *"For me, it's important to reflect or explain to them what's important, then to prioritize what's important, where their greatest weaknesses are, and to convince them to implement what we recommend. For me, these are the three most important points. (...) Past victims are generally more willing to implement measures than someone who just basically wants information and wants to find out a bit."* (I-B13). Indeed, advisors do inquire about their advisees' priorities. By contrasting desires like a new car or a lengthy vacation with the importance of burglary prevention, advisors can gain a clearer understanding of what the advisees truly value.

*Expertise state:* One's expertise state is the degree to which one has the knowledge, skills, and ability to diagnose one's own vulnerabilities, conceive sound solutions, and implement them or to teach others how to implement them. An expertise state ranges from having complete domain knowledge, skills, and abilities to having no domain knowledge, skills, and abilities.

Effective burglary prevention hinges on a grasp of technical and mechanical concepts. These can prove difficult for those without technical knowledge. In these cases, security advisors tailor their explanations to the individual's comprehension level. Some individuals even utilize their own technical or mechanical knowledge to propose creative security solutions. An advisor refers to their experiences with such advisees whom he calls *tinkerers*: *"I have also sometimes given services where you meet real tinkerers and hobbyists. They have already come up with the most amazing protection devices for their windows or cellar doors. That is in principle already very good. I do not say then: 'no, that does not go' but much more 'yes, that is almost perfect'"*. (S-B06). This advisor's receptiveness to unconventional and uncertified solutions not only accommodates advisees' technical understanding but also encourages their motivation to address security concerns.

*Trust state:* A trust state is the degree to which one is certain that the persuader is benevolent, competent, and attentive to one's vested interests and concerns. A trust state can range from full certainty in another's benevolence, competence, and attentiveness to full confidence that these qualities are absent.

Advisees approach their advisors with varying degrees of trust. In one observed advisory session, a couple displayed a noticeable difference in their perception of the police as an institution, and consequently, their advisor. The husband expressed skepticism towards each recommendation, while the wife remained overly enthusiastic about every suggestion. According to the advisors, not all advisees trust them or are open to the advice at all: *"I have sometimes had customers who have only become involved in a conversation over time. I could tell them what I wanted, but they were really angry or more like 'know-it-alls' at the beginning. But if I present the technical knowledge professionally and at the same time wittily, then I can also have a certain influence on them."* (S-B07). All advisors consistently employed the behavior of establishing their expertise in conversations through knowledge, rather than relying on their institutional power. They frequently achieved this by referencing insider information, such as technical details from previous burglary cases, aiming to make a favorable impression on the advisees.

*Reasoning state:* A reasoning state is the degree to which one is using critical thinking vs. ideological thinking. In critical thinking, one assumes one's position is provisional, and seeks evidence and logic to disconfirm it. In ideological thinking, one makes a subconscious assumption that one's position is truth, so one evaluates the validity of evidence and logic by the degree to which it is with it confirms one's position.

The advisors encounter individuals who hold diverse beliefs about burglars, many of which stem from mass media. These beliefs can be so deeply ingrained that advisors find it difficult to address them effectively. They have various methods to deal with counterproductive pre-assumptions as one of them illustrates: *"My opening sentence is usually this: 'Forget everything you see on TV now. And now just believe everything I'm about to tell you. First, it's sound, second, it's my job, and third, I know better than all the directors who write any movies.' It's quite funny sometimes. And sometimes it's difficult, because especially… Well, especially explicitly the older ladies, who are worried at home, maybe live alone, of course think they fall victim to a violent crime at home, because they watch 'File XY'. I say, 'No, you're not falling victim to a violent crime at home because that's statistically such a small chance, then you really don't have to worry.' In other words, that's where reassurance is called for, and that's where it's simply called for, for a little bit, to use the statistics, to simply say in no uncertain terms."* (I-B10). In this excerpt, the advisor describes two strategies. They first establish competence by drawing a parallel with movie directors. Next, they leverage statistics to highlight key facts. This tactic of using statistics is frequently employed to prioritize risks and persuade advisees.

## 5.2. Perceived Traits

Exploration of persuadees' states revealed that persuaders need to diagnose and respond to a persuadee's traits in real time too. A trait is a relatively stable, consistent, and enduring quality or characteristic that distinguishes one person from another. While states tend to fluctuate in time, traits typically change slowly over extended periods, if at all.

*Emotional stability:* The emotional stability describes the degree to which one experiences positive vs. negative emotions. People at the stable end of the continuum tend to be optimistic and emotionally resilient. They tend to be relaxed under normal conditions and calm in stressful situations. People at the non-stable end of the continuum (neuroticism) tend to be insecure and worried. They tend to experience negative emotions, even under normal conditions, and tend not to bounce back quickly when things go wrong. They are prone to mood swings, and tend to be irritable, moody, and sad.

The advisors encounter advisees who have different levels of sensitivity to external stimuli. The advisors claim they need to recognize persons who might develop strong emotions during an encounter such that they can keep up the professional character of the encounter and do not traumatize highly sensitive individuals. An advisor says: *"That is a matter of consideration. (...) You always have to look at who you're talking to. Are they sensitive? (...) Then you actually have to leave out pictures and videos, because some advisees are disturbed enough, and you also have to be more careful with the choice of words, which is quite odd."* (I-B13). The advisor notes that some of the tools, such as videos and pictures, could evoke strong emotional responses in certain advisees. Also, careful word choice is also essential to keep advisees stable.

*Agreeableness:* The agreeableness describes the degree to which one tends to put the needs of others before their own. People with high agreeableness tend to be altruistic, empathetic, friendly, helpful, polite, modest, cooperative, and compassionate. They may also find it difficult to assert their own needs. People with low agreeableness (disagreeableness) tend to be assertive, competitive, selfish, uncompassionate, and unhelpful. They are perceived as abrasive, critical, argumentative, stubborn, manipulative, and suspicious.

Advisors distinguish between interactions with advisees who are open to their recommendations and those who dismiss them outright, refusing even to collaborate on finding an alternative. Advisors emphasize that encounters with the latter are highly demotivating and unpleasant. Consequently, some advisors significantly reduce their efforts to persuade when faced with such resistance. An advisor puts it this way: *"There is one third [of advisees of whom] I am of the opinion that they do nothing. But you notice that right at the beginning of the conversation. (...) You notice it in the first five minutes, they are actually resistant to advice. (...) They know everything better. They always have a different opinion about everything and such consultations with me... so I usually cut them short. I don't go around the house with them for an hour and a half or so and go through everything, but I can do it in 15 minutes."* (I-B13). This advisor describes a radical approach for handling disagreeable advisees: significantly reducing persuasion attempts. While not the typical response, this highlights the significant challenge advisors face when dealing with disagreeable advisees. Such challenges can even lead to resignation in some cases, when advisors are repeatedly confronted by difficult individuals.

*Cognitive Capacity:* Cognitive capacity is described by the rate at which one stores and fetches frames from memory. It can range from very low, where a persuadee has few memory slots, less-elaborate frames, and/or slow storage and retrieval, to very high, where a persuadee has fast fetch and retrieval to and from working memory, many working memory slots, and highly elaborated frames.

In our data, some persuadees tended to respond to a persuader's utterances with no lag, while others tended to inject filler words or brief silences, which may indicate differences of store and fetch speeds. Persuaders had to adapt their pace to the speed at which persuadees could assimilate the utterances. Advisors are mindful of tailoring their approach when communicating with individuals who have varying levels of comprehension. They recognize that some people may require more support and explanation to grasp the details of a burglary case, while others are faster at that. An advisor describes his approach: *"The most important thing is when they are asking questions. You can tell by the questions, whether they've grasped what you've shown or what you've explained, you can tell by the questions. (...) Depending on what kind of technical questions they ask. If you explained it on a normal level, you would notice, he understood it, depending on the technical questions, you notice, he understood it super or, now there is still a gap, now I have to go into it more, regarding that."* (I-B18). This advisor attends to questions asked by the advisee as a proxy to assess their cognitive capacities.

## 6. Discussion

This retrospective analysis examines the factors burglary prevention advisors consider during persuasive interactions. While advisors' approaches may vary slightly, their underlying concerns remain consistent across the data set. For example, when faced with a distressed and sensitive individual, one advisor might emphasize statistics to alleviate anxiety, while another might opt for a more empathetic and emotionally-driven communication style. Nevertheless, this emphasis on the advisee's emotional state was a recurring theme in both our observations and the advisors' own reflections.

Our analysis suggests an answer to the question of what drives idiosyncratic behavior in advisors. We find that advisors' behaviors shift based on their assessments of their advisees' states and traits. Their persuasive efforts are shaped by their perceptions of the advisees' desires and needs. Contrary to previous research emphasizing advisors' institutional identity (Drew & Heritage, 1992; Svinhufvud & Vehviläinen, 2013), limitations of their tools (Dolata et al., 2020), impression management (Dolata & Schwabe, 2017), or individual preferences (Dolata & Schwabe, 2018) as primary drivers, our study reveals a nuanced picture. Advisor behavior emerges from a complex process of quickly and accurately assessing the person they are advising and adapting their behavior. This involves navigating a balance between the stereotypes associated with their institutional role—which, in the case of police advisors, is closely tied to their authority and public perception—and the need for personalized interaction.

This implicit navigation processes became apparent through the advisors' post-hoc reflections quoted above. Notably, during the advisory sessions, they maintained focus and confidence, acting without explicitly addressing concerns about the advisee's state or characteristics, yet the changes in their behavior could be identified in the documented observations. This suggests that their adaptations are intuitive, likely stemming from experience rather than formal training. This insight might also explain the inconsistent use of some tools provided by the SP system, despite being co-designed with the advisors. Instead of focusing on a single advisee profile, the research team and advisors should have considered diverse advisee profiles during the design phase. This broader perspective would have better equipped advisors to navigate the nuances of interpersonal dynamics inherent in advisory services, leading to more effective tool integration (Dolata et al., 2019). Further research into user profiles is crucial, particularly in contexts involving collaboration between two or more individuals.

However, the advisors' heavy reliance on personal assessments, often formed within "the first five minutes" as one advisor admitted, could introduce bias. Rapid judgments often stem from personal stereotypes based on experience. Yet, society is constantly evolving, and so are the advisees. Advisors should be made aware of their own biases and assumptions, learning to leverage them when appropriate, but also to set them aside when necessary. Technology can play a role in this process. For example, it could dynamically alert advisors to potential errors stemming from stereotypes, helping them to provide more objective and relevant guidance.

Our findings reveal which advisee characteristics are essential for advisors who tailor persuasive communication. This study reinforces previous research by highlighting the continued relevance of factors such as emotional state or mood (Oyebode et al., 2024; Stavraki et al., 2021; Vaz & Sousa, 2021) and resistance to advice, which relates to agreeableness and trust (Beutler et al., 2018; Krebs et al., 2018). Additionally, our data suggests that the "reasoning state" corresponds to the previously identified, yet under-researched, aspect of counterfactual thinking (O'Keefe, 2016). Overall, these identified states and traits align with existing literature.

Our findings highlight additional variables that advisors consider during persuasive interactions. Specifically, emotional stability, perceived expertise, self-efficacy, cognitive capacity, and risk aversion emerged as significant factors. We posit that the importance of these aspects stems from the nature of advice itself.

For example, cognitive capacity and perceived expertise become particularly relevant due to the complex, technical nature of burglary prevention strategies. Risk aversion and emotional stability are intertwined with emotions such as fear, anxiety, and feelings of loss or vulnerability—all common reactions to the topic of violence, crime, and burglary. Self-efficacy, on the other hand, relates to an individual's belief in their own abilities. While the concepts of ability and empowerment are well-established in persuasion research (Fogg, 2009; Oinas-Kukkonen & Harjumaa, 2009; Stibe, 2015), self-efficacy has received less attention. Advisors acknowledge that they can only influence a person's actual ability to a limited extent, especially when external factors like financial resources are involved. Yet, they can impact how individuals perceive their ability to enhance safety by suggesting more accessible and affordable solutions. Overall, this study identifies key aspects to consider when tailoring persuasive efforts to the individual characteristics of the recipient.

The close relationship between some emerging variables and the specific, dynamic context further challenges the "one-size-fits-all" approach prevalent in PT literature. It supports previous research advocating for personalization (Aldenaini et al., 2020; Alslaity et al., 2024; Kaptein et al., 2015; Orji & Moffatt, 2018). Additionally, it also highlights the need for more domain-specific persuasion research. Expertise, for instance, might be crucial in complex situations requiring multi-layered intervention. However, it might be less impactful for simpler advice, such as persuading the patient to regularly taking medication. This could explain why some variables identified in this study are less discussed in literature on personalized persuasive systems within the medical field (Alslaity et al., 2024). Similarly, emotional stability might be more critical in domains involving potential risks or injuries, while its impact could be low in situations where health and safety are not a concern. While our findings could benefit related fields impacted by the proliferation of persuasion and collaboration technologies like weight-loss programs, psychotherapeutic support for addiction management, and energy-saving interventions, the importance of persuader variables might differ for advisors and advisees in those contexts. Therefore, further research is needed to determine the applicability of our results to these areas.

Furthermore, advisors emphasize the dynamic nature of interactions during service delivery. Persuasive systems, therefore, might need to incorporate dynamic variables and account for their evolution over time. For example, a person's emotional state, like fear, can shift during a persuasive conversation, potentially influenced by new information. This necessitates system adaptation, a task easily performed by human advisors but still challenging to replicate technologically. The ability of technology to dynamically understand and respond to a user's evolving state remains a significant hurdle.

Current research is exploring the use of generative AI for persuasion (Matz et al., 2024; Wu et al., 2024). This trend is likely to intensify, leading to the prolifera-tion of persuasive agents in healthcare or fitness. Generative AI offers new personalization methods, allowing systems to reference user interactions or behaviors directly. Yet, these systems require careful calibration through prompting or training. It is crucial for them to consider factors like a user's risk aversion, emotional stability, and self-efficacy when employing persuasive techniques. This will help mitigate the risk of generating potentially harmful messages, such as overly intense content for someone with low emotional stability. Also, combining automatically generated persuasive messages with the oversight of experienced human persuaders could enhance the quality of persuasion overall by incorporating their expertise. This collaboration could lead to more flexible and effective AI-based persuasive agents who can learn from the human persuaders.

In human-humn, such as SP, generative AI could personalize multimedia and messages. There is an obvious way to use the identified states and traits for design: A generative AI-based agent may be prompted to establish the traits of the persuadee in the needs elicitation phase and behave adequately in the subsequent phase. Instead of omitting information, advisors could utilize AI to create content that resonates with the advisee and integrate it into the interaction. For example, persuadees with high emotional stability and low risk aversion might respond well to footage of actual burglaries, while others might be more receptive to technical explanations or visuals demonstrating how security measures prevent intrusions. AI could further assist by personalizing documentation, streamlining solution selection, and addressing identified security vulnerabilities.

This manuscript offers valuable insights into the dynamics of persuasive encounters, carrying significant implications for designing and developing persuasive systems. By adopting the persuader's perspective, the authors identify potentially relevant personalization variables, thereby expanding the existing discourse on personalization within persuasive systems. We introduce a distinction between states and traits, differentiating between more dynamic and stable aspects requiring consideration. Finally, the manuscript illustrates how advisors navigate these aspects, indicating how potential persuasive systems could handle similar situations.

There is significant potential for research to delve deeper into the nuances of persuasion involving both human persuaders and technology. The field should move beyond the computer-human influence (Oinas-Kukkonen & Harjumaa, 2009; Stibe, 2015), as individuals are increasingly exposed to persuasive situations where efforts are divided between human and technological agents. One example is a doctor recommending behavioral changes, such as improved eating habits, with technology subsequently reinforcing these recommendations through reminders (Staehelin et al., 2023). Further research is crucial to understand how such combined approaches compare to those relying solely on technology or human interaction. We posit that these combined approaches, particularly when personalized, may prove more effective than their single-agent counterparts.

## 7. Conclusion

This manuscript draws on data from an extensive project on burglary prevention support to examine how individual attributes of the persuadees impact the persuaders' behavior and use of additional tools or persuasion techniques. We acknowledge certain limitations inherent in our study design. The study's reliance on a large dataset carries a potential risk of selectively choosing data to fit a narrative. To mitigate this, we involved three researchers in the analysis and interpretation process, ensuring multiple perspectives were considered. Additionally, the retrospective nature of the study presents a risk of relying on memory over objective data. To address this, we compiled a comprehensive collection of critical incidents and quotes prior to constructing the narrative presented in this paper. We acknowledge that the identified advisee factors (like risk aversion or emotional stability) are intertwined with the specific context of burglary prevention advice and based on advisor's statements. Future research incorporating findings from other fields and independent measures could bolster our claims, particularly for agreeableness and reasoning, likely influential across contexts. Finally, while rooted in DSR, this study lacks specific design guidance. Although our prior research (Comes & Schwabe, 2016a, 2016b) reveals ways to enhance persuasiveness, future work should explore the identified variables and deliver designs to support personalization.

The findings presented in this manuscript is relevant for researchers studying advisory services, offering insights into the idiosyncratic practices they encounter. Additionally, this work provides valuable considerations for designing and researching persuasive systems, particularly for those interested in personalization. By highlighting the persuadee as an active participant rather than a passive recipient, this research encourages persuasion scholars to adopt a nuanced perspective on the persuasion process. Practitioners in advisory services gain an understanding of the factors that influence advising, while engineers gain valuable insights into the advisors' behaviors and the motivations behind them.